\documentclass{aa}
\include{psfig}

\def\e20{\times 10^{20} {\rm cm}^{-2} }
\def\ecs{\rm \,erg~ cm^{-2}\,s^{-1} }
\def\ergs{\rm \,erg~ s^{-1} }
\def\aro{\alpha_{\rm ro} }
\def\arx{\alpha_{\rm rx} }
\def\aox{\alpha_{\rm ox} }

\begin{document}
   \thesaurus{03     
              (11.01.2;  
               11.05.2;  
               11.12.2;  
               11.17.3;  
               13.18.1;  
               13.25.2)} 

\title{An X--ray selected sample of radio--loud quasars}
\subtitle{  }
\author{Anna Wolter  \inst{1}  \and Annalisa Celotti \inst{2}}
\offprints{Anna Wolter, {\it anna@brera.mi.astro.it} }
\institute{Osservatorio Astronomico di Brera, via Brera 28, I-20121
Milano, Italy email: anna@brera.mi.astro.it \and S.I.S.S.A., via
Beirut 2-4, I-34014 Trieste, Italy email: celotti@sissa.it}
   
\date{Received 29 August 2000; accepted ...}

\maketitle
\begin{abstract}
We construct the first X-ray selected sample of broad line radio-loud
AGN from the EMSS survey. In order to test unifying schemes for
radio-loud objects, their spectral and statistical properties (both
flat and steep spectrum objects) are examined and compared with those
of other samples of blazars.  The X--ray selection allows us to
explore properties of radio--loud quasars 10--100 weaker in the radio
band than classical samples.  The most convincing interpretation of
our results is that there are no radio--loud quasars whose synchrotron
emission peak reaches the EUV--soft X--ray band at these (radio) flux
levels.  Moreover, due to the comparatively weak non-thermal emission,
a quasi--thermal component appears to contribute at optical--UV
energies.  The detection of sources at low radio fluxes also reveals a
large population of steep spectrum quasars and the lack of the
predicted turnover in the quasar (radio) counts. The evolution of
X--ray selected radio-loud quasars does not significantly differ from
that of radio--selected ones, and flat spectrum and steep spectrum
sources appear to behave quite similarly.  

\keywords{Galaxies: active
-- Galaxies: evolution -- Galaxies: luminosity function -- Quasars:
general -- Radio continuum: galaxies -- X-rays: galaxies} 

\end{abstract}


\section{Introduction}

The selection energy band in the construction of complete samples
constitutes one of the main biases and a source of confusion in the
understanding of the Active Galactic Nuclei (AGN) phenomenology, which
has often led to useful but somehow unphysical classifications.

Among radio--loud objects, a specific and instructive example -- which
will constitute a reference in the following -- is given by the case
of BL Lac objects. Samples of BL Lacs selected in the radio and X--ray
bands revealed systematic differences in their spectral energy
distributions (SED), polarization, variability properties and
evolution (e.g. Urry \& Padovani 1995).  Different models have been
proposed to account for this apparent ``dichotomy'' (Stocke et
al. 1985; Ghisellini \& Maraschi 1989; Celotti et al. 1993), and only
recently it has been explicitly recognized that an intrinsic
continuous distribution in the position of the peak of the broad band
SED of these sources might naturally lead to the detection of
different types of objects depending on the selection band (Padovani
\& Giommi 1995). This new perspective has allowed us to have a
better view of the intrinsic, physical properties of BL
Lacs and speculate on their origin.

The SED of all blazars can be well represented by two main spectral
components, believed to be produced as synchrotron and inverse Compton
emission, respectively.  One suggested possibility then is that the
broad band spectral distributions of `X--ray selected' (XBL) and
`radio--selected' (RBL) objects are simply distinguished by the energy
location of the peak of the synchrotron emission: at UV/X--ray in the
former ones (High peak BL Lacs, HBL) and around the optical in the
latter ones (Low peak BL Lacs, LBL; Padovani \& Giommi 1995), while
the Compton peak is located in the $\gamma$--ray band (from MeV to TeV
energies).

Furthermore, the entire blazar population, ranging from HBL, to LBL to
Flat Spectrum Radio Quasars (FSRQ) can be interpreted, so far, as a 
sequence from lineless, low
power sources, to broad lined, powerful quasars characterized by the
(decreasing) peak energies of the two spectral components, the
increasing source radio power and the increasing ratio between the
Compton and synchrotron luminosities (Fossati et al. 1998). This trend
in the SED can be physically accounted for by an increase in an
external soft photon field along the sequence, in agreement with the
observed behavior (Ghisellini et al. 1998). Clearly, this scenario
makes definite predictions on the relationship between (radio) source
power and spectral shape.

In order to understand the properties of a population it appears
therefore crucial to select sources in different spectral bands, and
compare their characteristics. To this aim, i.e. disentangling the
intrinsic features from the selection induced ones, we have considered
radio--loud quasars -- which so far have been extracted from radio
surveys -- and studied their broad band and cosmological properties
when selected in X--rays.

The sample comprises the radio--loud quasars (i.e. broad line 
AGN) detected in the {\it Einstein} Medium Sensitivity
Survey (EMSS; Gioia et al. 1990, Stocke et al. 1991). More
specifically, we have constructed the first statistically complete
X--ray selected sample of (39) radio--loud quasars, in order to: a)
determine whether their spectral properties differ from those of the
radio--selected quasars and other classes of radio--loud objects; b)
discuss the properties of these objects in the context of the beaming
scenario and emission mechanisms for blazars; c) determine the
statistical properties of this new sample, in the light of the
unification scheme for powerful radio--loud objects (e.g. Urry \&
Padovani 1995 and references therein), where flat spectrum radio
quasars are high power (FR~II) radio galaxies in which the
relativistically beamed component dominates.

With the selection criteria adopted, the sample comprises both
Flat Spectrum (FS, blazar--like) and Steep Spectrum (SS) radio 
quasars~\footnote{Note that the sample is not selected on the point--like
optical appearance, and thus the SS sources possibly include Broad
Line Radio Galaxies.}.  It should be reminded that the radio emission
in the latter ones is believed to originate from the extended radio
components.  In these sense the SS quasars will be not directly
included in the comparison of blazar properties, but we will study and
examine their optical, X--ray spectral characteristics and their
statistical properties with respect to the FS quasars in this sample.

After discussing the sample selection (Sect. 2), we present the
broad band spectral properties of the objects, distinguishing between
flat and steep radio sources (Sect. 3) and compare them with those
of other blazars (Sect. 4).  The statistical and evolutionary
properties of the new sample are derived in Sect. 5, while in the
final Sect. 6 we discuss our results.

Hereafter we adopt cosmological parameters $H_{\rm 0}=50$ km s$^{-1}$
Mpc$^{-1}$ and $q_0=0$.


\section{The sample}

The EMSS provides us with well defined X-ray selected samples of
extragalactic sources. The virtually complete identification status
makes it a very powerful tool to study statistical properties of
classes and sub-classes of astronomical sources.

Radio information for all the (extragalactic) EMSS objects is
available from VLA snapshots at 6 cm taken during the identification
process (e.g. Maccacaro et al. 1994), with a flux limit of about 1
mJy, corresponding to 5 times the rms of the observations.  Since we
intend to compute the radio spectrum using the recently released data
at 20cm from the NRAO VLA Sky Survey (NVSS; Condon et al. 1998), we
cut the EMSS sample at $\delta \geq -40^{\circ}$.  With this cut, that
excludes only 5\% of all the AGNs and only one radio--loud, the EMSS
contains 79 radio detected AGN, of which 43 have a two-point spectral
index\footnote{ While the original EMSS two point spectral indices
$\aro$, $\aox$, $\arx$ were estimated at 5 GHz, 2500 \AA\, and 2 keV,
comoving energies, here we recompute them at 5 GHz, 5500 \AA\, and 1
keV, comoving energies, in order to compare with values for other
samples. For the K-correction in the optical and X--ray bands it is
assumed $\alpha_{\rm o}= \alpha_{\rm x} =1$ (see e.g. Della Ceca et
al. 1994), where the spectral indices $\alpha$ are defined by $F_{\nu}
\sim \nu^{-\alpha}$.} between radio and optical ($\aro$) larger than
0.35 (Della Ceca et al. 1994) and are therefore classified as
radio--loud. We exclude another four objects for the reasons detailed
in the Appendix, thus constructing a complete sample of 39 radio-loud
quasars, that we dub the EMSS Radio-Loud quasar sample
(ERL).

In the following we will refer to FS and SS spectrum radio--loud
objects, where the division is formally considered to be at
$\alpha_{\rm r}$=0.7 (see e.g. Perlman et al. 1998) which leads to a
similar statistics for the two sub--classes (20 FS and 19 SS). As the
distribution of $\alpha_{\rm r}$ does not show any suggestion of being
the sum of two different (bimodal) distributions (see
Fig.~\ref{distr}), the choice of a dividing value is somewhat
arbitrary.  In any case, the results of this work are not affected by
a different choice (the case $\alpha_{\rm r}$=0.5 has been also
explored).

The list of sources and information is presented in Table~\ref{lista}.
Columns are as follows: (1) Name of the source in the EMSS catalog --
numbers refer to notes to the Table, indicating either other names
from literature or the source radio morphology; (2) Redshift; (3) 5
GHz flux from EMSS, unless otherwise stated, in mJy; (4) 1.4 GHz flux
from NVSS; (5) Radio spectral index; (6) Magnitude, from EMSS; (7)
unabsorbed X-ray flux (0.3-3.5 keV band), from EMSS; (8) Galactic
$N_{\rm H}$, from Stark et al. (1992); (9) X-ray spectral index from
EMSS data; (10) X-ray spectral index from ROSAT data (Brinkmann et
al. 1997); (11-13) $\aro$, $\aox$ and $\arx$.

The distributions of radio spectral indices and luminosities in the
three bands are plotted in Fig.~\ref{distr} for FS and SS,
respectively.

\section{Spectral properties}
\subsection{X-ray spectral indices}

The X--ray spectral indices for the sample have been estimated from
the IPC hardness ratios (see e.g. Maccacaro et al. 1988), assuming a
power law spectral shape and Galactic line of sight absorbing columns,
and compared with those determined by ROSAT.  The former ones span a
large interval $-2.2< \alpha_{\rm x} < 2.4$, and within the
uncertainties significantly correlate with the ROSAT values (see
Fig.~\ref{figax}). Unfortunately the large errors do not allow either
to detect any difference between FS and SS sources, or to infer the
origin of the X--ray component (e.g. steep synchrotron dominated
versus flat Compton dominated emission).

No relationship has been found between the X--ray and radio spectral
indices, the source fluxes or redshifts.

\begin{table*}
\caption{The sample}
\begin{tabular}[h]{| l r r r r r r r r r r r r|}
\hline
Name &  z& F$_{5}^{\rm a}$ & F$_{1.4}^{\rm b}$ &$\alpha_{\rm r}$& mag &$F_{\rm X}^{\rm c}$ & $N_{\rm H}^{\rm d}$ & $\alpha_{\rm X}$  & $\alpha_{\rm ROSAT}$   & $\aro$& $\aox$ &$\arx$ \\
     & &              &                &          &     &    &  & &    & & & \\
 MS0012.5$-$0024&         1.695&   24.0&      14.9& $-$0.40&18.51  &    2.41&  2.9&  1.50$^{+0.70}_{-0.63}$&                  & 0.32&  1.33&  0.67\\
 MS0038.8$-$0159$^1$&     1.675&  322.0&    1034.6&    0.97&16.86  &    3.07&  2.9& --0.31$^{+0.69}_{-1.17}$&                  & 0.53&  1.54&  0.88\\
 MS0039.2$-$0206&         1.083&    5.6&       7.0&    0.19&19.00  &    1.48&  2.9&  ---                   &                  & 0.30&  1.33&  0.66\\
 MS0053.3$-$1035&         0.308&   27.6&      40.6&    0.32&18.44  &    3.33&  3.7&  ---                   &                  & 0.42&  1.29&  0.72\\
 MS0136.3+0606&           0.450&   16.3&      67.5&    1.18&18.52  &    2.72&  3.7&  2.07$^{+1.24}_{-0.78}$&   1.78$\pm$0.16  & 0.41&  1.30&  0.72\\
 MS0152.4+0424&           1.132&  184.1&     323.8&    0.47&19.17  &    6.28&  4.2&  0.50$^{+0.50}_{-0.63}$&                  & 0.63&  1.07&  0.78\\
 MS0225.5$-$1052&         1.038&   60.0&     132.2&    0.66&19.34  &    1.56&  2.5& --0.66$^{+1.66}_{-\infty}$& 1.08$\pm$0.15  & 0.56&  1.29&  0.81\\
 MS0226.8$-$1041$^2$&     0.620&   96.2&     293.2&    0.93&18.32  &    2.53&  2.5&  1.13$^{+0.47}_{-0.48}$&   0.92$\pm$0.19  & 0.54&  1.35&  0.82\\
 MS0232.5$-$0414$^3$&     1.450&  522.9&    1495.9&    0.87&16.28  &    7.19&  2.5&  1.15$^{+0.38}_{-0.39}$&   1.18$\pm$0.06  & 0.51&  1.49&  0.85\\
 MS0311.8$-$0801&         1.250&    8.2&      70.3&    1.78&19.12  &    3.38&  5.4&  1.10$^{+0.92}_{-1.19}$&                  & 0.44&  1.17&  0.70\\
 MS0402.0$-$3613$^4$&     1.417& 2019.0&    1132.3& $-$0.48&17.17  &   33.98&  0.8&  0.59$^{+0.13}_{-0.13}$&   0.75$\pm$0.16  & 0.59&  1.12&  0.78\\
 MS0438.6$-$1050&         0.868&   26.8&      51.2&    0.54&18.97  &    3.48&  5.9& --1.45$^{+1.90}_{-\infty}$&                & 0.46&  1.19&  0.71\\
 MS0521.7+7918&           0.503&   22.1&     102.6&    1.28&17.45  &    5.52&  7.9&  0.98$^{+1.41}_{-2.20}$&                  & 0.35&  1.33&  0.69\\
 MS0808.0+4840&           0.700&   75.0&     160.3&    0.63&17.96  &    2.97&  4.6&  1.45$^{+0.80}_{-0.85}$&   1.32$\pm$0.16  & 0.48&  1.36&  0.78\\
 MS0815.7+5233$^5$&       0.624&    1.5&       3.6&    0.73&20.75  &    3.15&  4.3&  2.03$^{+0.93}_{-0.72}$&                  & 0.37&  0.94&  0.56\\
 MS0822.0+0309&           0.577&   31.5&      85.3&    0.83&17.94  &    1.46&  3.5&  2.85$^{+9.99}_{-1.78}$&                  & 0.41&  1.49&  0.78\\
 MS0828.7+6601&           0.329&   98.8&     235.3&    0.72&20.21  &    4.90&  4.2& --0.08$^{+0.53}_{-0.71}$&                  & 0.69&  0.95&  0.78\\
 MS0833.3+6523$^6$&       1.112& 369.0$^{\rm e}$& 1275.7& 1.03&18.16 &  3.84&  4.4& --0.54$^{+1.00}_{-2.33}$&   0.46$\pm$0.24  & 0.65&  1.29&  0.87\\
 MS0849.0+2845&           1.273&  336.0&     232.6& $-$0.31&19.09  &    1.73&  3.2& ---                   &                & 0.62&  1.29&  0.85  \\
 MS0850.2+2825&           0.922&   37.0&      68.0&    0.51&18.27  &    1.37&  3.2& --2.48$^{+2.33}_{-\infty}$&                & 0.42&  1.46&  0.78\\
 MS0958.4+6913$^7$&       0.928&   23.0&      59.8&    0.79&19.08  &    2.68&  3.7& --0.24$^{+0.67}_{-1.04}$&   0.24$\pm$0.18  & 0.47&  1.22&  0.73\\
 MS1003.6+1300$^{2,8}$&   0.648&  125.0&     305.7&    0.74&17.45  &    2.75&  3.8&  1.38$^{+0.68}_{-0.70}$&                  & 0.49&  1.46&  0.82\\
 MS1050.9+5418&           0.995&  167.1&     266.0&    0.39&19.25  &    4.89&  0.8&  0.42$^{+0.49}_{-0.59}$&  -0.02$\pm$0.45  & 0.63&  1.12&  0.79\\
 MS1138.6+6553$^2$&       0.805&  132.8&     506.4&    1.11&18.30  &    1.26&  1.0&  0.75$^{+0.53}_{-0.56}$&   1.00$\pm$0.27  & 0.57&  1.48&  0.89\\
 MS1234.9+6651$^{2}$&    0.860&  142.2&     335.2&    0.71&17.99  &    5.39&  1.9& --1.21$^{+1.71}_{-\infty}$&                & 0.53&  1.29&  0.79\\
 MS1311.1+3210$^{9}$&    0.303&    7.0&     132.0&    2.44&18.86  &    3.29&  1.1& --0.31$^{+1.09}_{-\infty}$&                & 0.39&  1.24&  0.69\\
 MS1326.6+2546$^{2,9}$&    0.986&   45.6&      89.3&    0.56&17.80  &    2.40&  1.1&  0.76$^{+0.47}_{-0.50}$&   0.60$\pm$1.00  & 0.41&  1.45&  0.77\\
 MS1340.7+2859&           0.905&  250.0&     252.1&    0.01&17.05  &    4.88&  1.2&  0.55$^{+0.98}_{-1.41}$&   1.06$\pm$0.38  & 0.47&  1.44&  0.80\\
 MS1442.8+6344&           1.380&  445.0&     689.6&    0.36&17.19  &    2.93&  1.7&  1.05$^{+0.47}_{-0.47}$&   1.12$\pm$0.16  & 0.53&  1.51&  0.87\\
 MS1623.4+2712$^{10}$&    0.525& 240.0$^{\rm f}$& 531.0& 0.66&18.41&    7.78&  3.4&  1.38$^{+0.32}_{-0.32}$&   1.96$\pm$0.39  & 0.62&  1.16&  0.80\\
 MS1640.0+3940&           0.540&   31.0&      54.4&    0.47&18.25  &    4.39&  1.0&  0.93$^{+0.41}_{-0.39}$&   1.03$\pm$0.05  & 0.41&  1.29&  0.72\\
 MS1657.1+3524&           0.949&    5.6&       5.2& $-$0.06&20.24  &    1.11&  1.7&  0.62$^{+1.01}_{-1.49}$&   1.30$\pm$0.11  & 0.39&  1.21&  0.67\\
 MS1837.8+4538$^{2,11}$&  0.958&  210.0&     801.4&    1.11&19.38  &    3.60&  5.9&  0.84$^{+1.79}_{-\infty}$&                & 0.70&  1.12&  0.84\\
 MS2037.3$-$0035$^{2}$&   0.609&   23.1&      73.1&    0.96&18.39  &    2.21&  6.6&  1.93$^{+0.68}_{-0.67}$&   1.30$\pm$0.43  & 0.42&  1.34&  0.74\\
 MS2113.8+0455&           1.001&  122.0&     316.8&    0.79&20.37  &    2.53&  6.5& --0.02$^{+1.25}_{-3.03}$&   0.39$\pm$0.00  & 0.71&  1.02&  0.82\\
 MS2141.2+1730$^{12}$&    0.213&  417.0&     651.8&    0.37&15.85  &    8.94&  8.0&  2.68$^{+1.52}_{-1.15}$&   1.44$\pm$0.23  & 0.46&  1.49&  0.81\\
 MS2143.2+1429&           1.387&   50.3&     226.2&    1.25&20.40  &    1.91&  7.0&  ---                   &                  & 0.67&  1.06&  0.80\\
 MS2247.8$-$0703&         1.710&   44.7&      82.4&    0.51&20.82  &    3.82&  3.6&  1.26$^{+0.59}_{-0.62}$&                  & 0.63&  0.90&  0.72\\
 MS2329.3$-$3827$^{13}$&  1.195&  670.0&     543.6& $-$0.17&16.20  &    3.08&  1.5& --1.05$^{+2.02}_{-\infty}$&                & 0.46&  1.65&  0.87\\

\hline
\end{tabular}
\begin{list}{}{}
\item $^{\rm a}$ Flux @ 5 GHz in mJy; $^{\rm b}$ Flux @ 1.4 GHz in
mJy; $^{\rm c}$ Unabsorbed flux in (0.3-3.5 keV) in $\times 10^{-13}
\ecs$; $^{\rm d}$ Hydrogen column density in $\e20$; $^{\rm e}$ from
Gregory \& Condon (1991); $^{\rm f}$ from Becker et al. (1991).  
Notes: (1) 4C02.04 (2) Triple radio source 
(3) PKS 0232$-$04, double radio source
(4) PKS 0402$-$362 (5) Weak-lined, see text (6) 3C 204 (7) Radio
core+extended (8) WE 1138+65 (9) WAT (10) B2.2 1623+27A
(11) OU+462 (12) OX 169 (13) PKS 2329$-$384, HB

\end{list}
\label{lista}
\end{table*}

\begin{figure}
\psfig{file=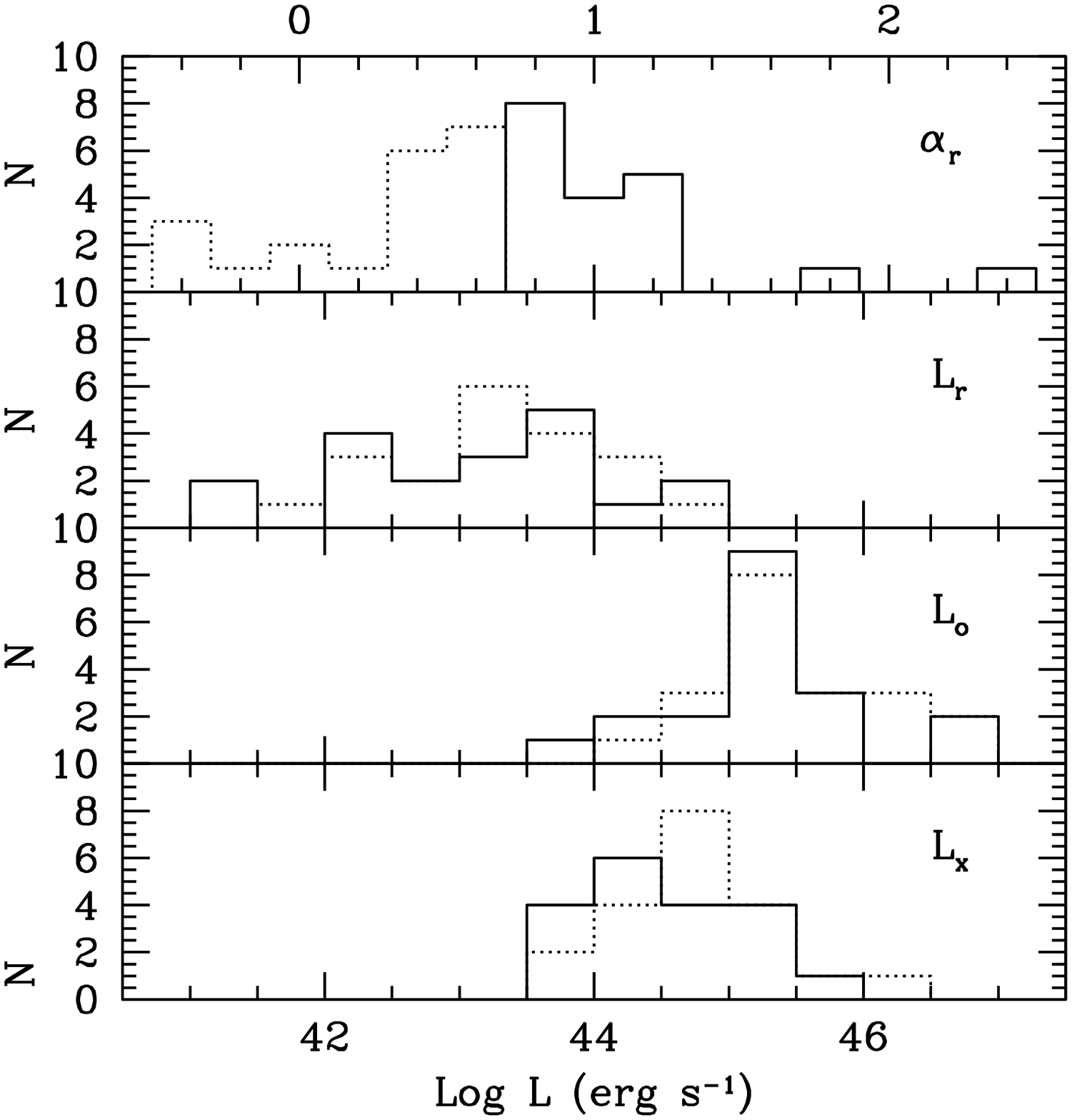,width=8truecm}
\caption{Distributions of radio spectral indices and luminosities for
the ERL. The FS and SS subsamples are indicated by the dotted
and continuous line, respectively.  The axis on top of the
figure refers to the spectral index $\alpha_{\rm r}$.}
\label{distr}
\end{figure}

\begin{figure}
\psfig{file=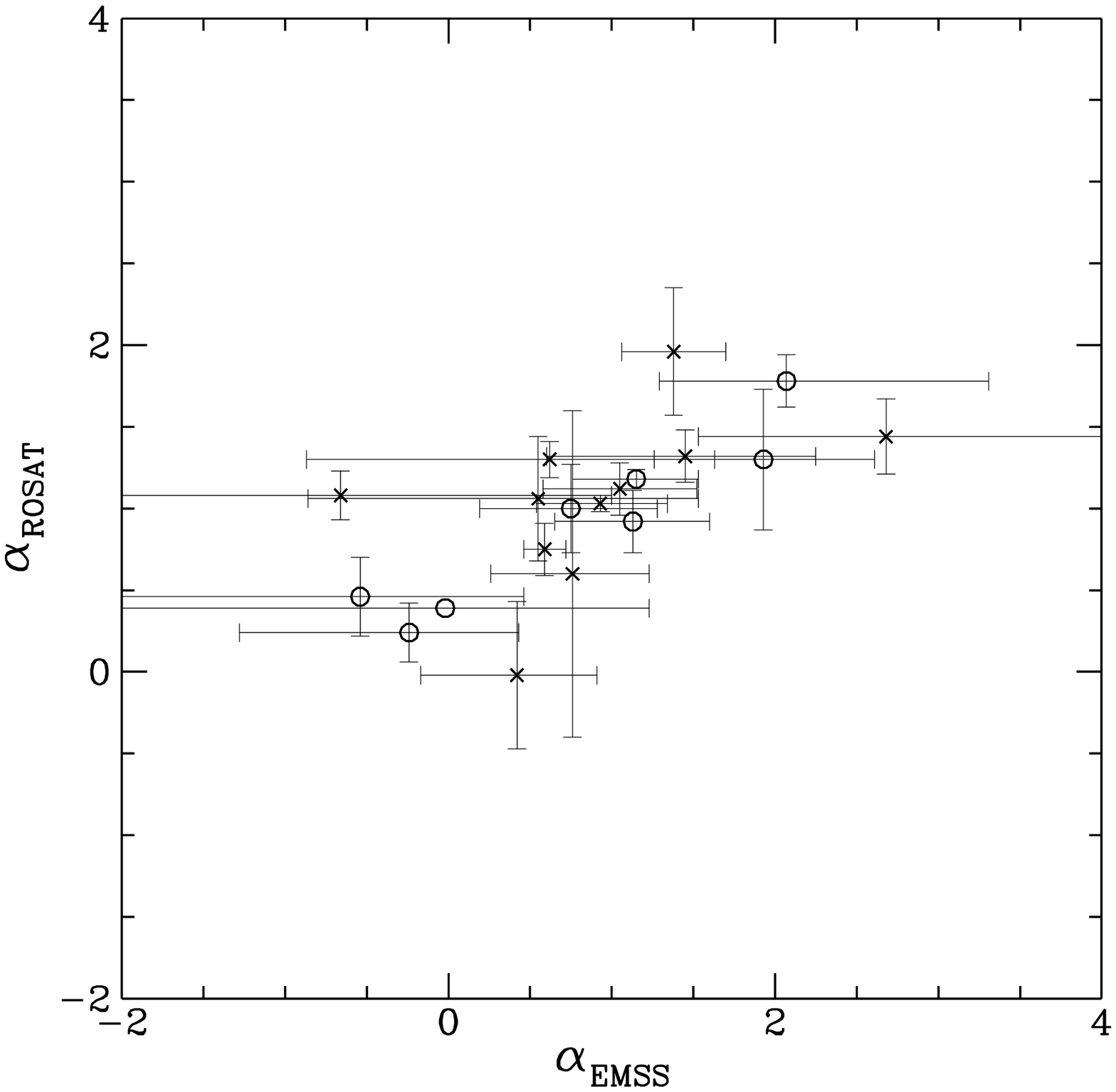,width=8truecm}
\caption{The ROSAT spectral index (from Brinkmann et al. 1997) vs. the
EMSS spectral index derived from the Hardness Ratio (from Maccacaro et
al.  1988); FS and SS sources are indicated as crosses and circles,
respectively. }
\label{figax}
\end{figure}

\subsection{Broad band properties}

In order to quantify the total power and SED properties of the sources
and compare them with those of other samples, we compute the
luminosities in the radio, optical and X--ray bands ($L_{\rm r}$,
$L_{\rm o}$, $L_{\rm x}$) and the corresponding broad band spectral
indices ($\aro$, $\aox$, $\arx$).

Significant linear correlations are found between the luminosities in
the three bands.  However, since a possible culprit for this kind of
correlation is the common redshift dependence, we test its influence
by using a partial correlation analysis (e.g. Kendall \& Stuart 1979).
The partial correlation coefficients are defined as:

\begin{equation}
r_{12\_3} = 
{{(r_{12} - r_{13} \times r_{23})} \over{\sqrt{(1-r_{13}^2) 
\times (1-r_{23}^2)}}}, 
\end{equation}
where $r_{12}$, $r_{13}$, and $r_{23}$ are the Spearman rank test
coefficients for two variables at a time.  By assigning to index ``3''
the redshift variable and interchanging $L_{\rm x}$, $L_{\rm r}$ and
$L_{\rm o}$ as variables ``1'' and ``2'' we obtain the following
results: $r_{rx\_z}$ = 0.441; $r_{ox\_z}$ = 0.325; $r_{ro\_z}$ =
0.381.  The null hypothesis is rejected at the 95\% level for the last
two occurrences, while the radio and X-ray luminosity seem to be
correlated at the 99\% level.

\begin{figure}
\psfig{file=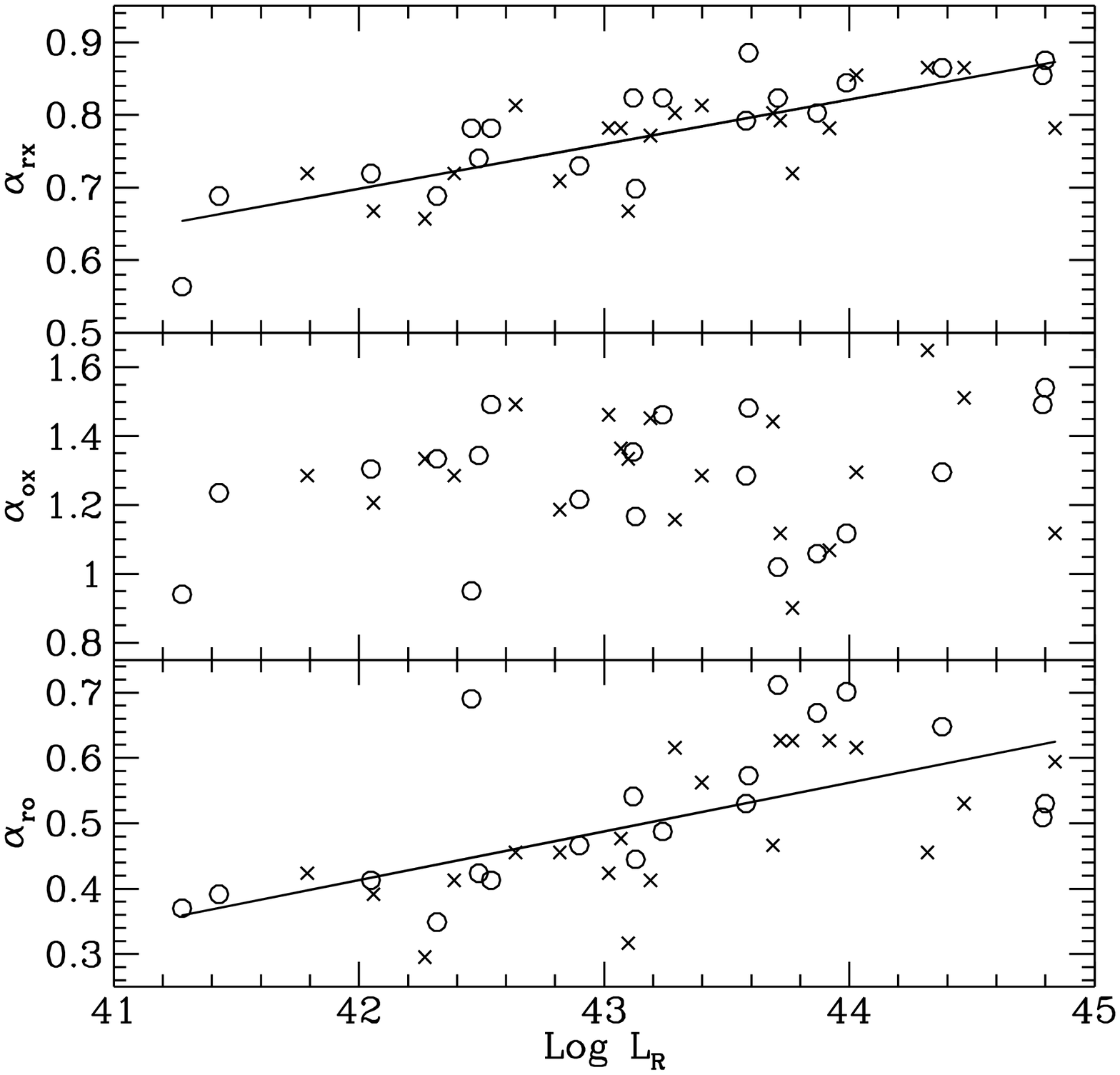,width=8truecm}
\caption{Distribution of $\arx$, $\aox$, $\aro$ as a function of
$L_{\rm r}$. SS are represented by circles and FS by crosses.  The
linear regression for the entire sample is plotted for $\arx$ and
$\aro$ (the separate regressions for FS and SS are consistent with the
total), while $\aox$ vs $L_{\rm r}$ is not statistically significant.}
\label{figalr}
\end{figure}

No differences (according to the Kolmogorov-Smirnoff test) are found
between the distributions relative to FS and SS, neither in
luminosities (probability $p\geq 20\%$; see Fig.~\ref{distr}) nor in
broad band spectral indices ($p \geq 47\%$; see Fig.~\ref{figalr}) nor
in the trends, suggesting a behavior of the compact beamed component
independent of the large scale radio flux dominance.

The shape of the SED appears however to be clearly related to the
radio luminosity: in fact, the only statistically significant trends
found are of both $\aro$ and $\arx$ with $L_{\rm r}$ (see
Fig.~\ref{figalr}).  Although the correlation might be due to the
common dependence of the two spectral indices on $L_{\rm r}$ itself,
its relevance is supported by two facts: a) $L_{\rm r}$ does not
correlate with $\aox$; b) $L_{\rm r}$ spans the largest interval in
luminosity (about five decades compared to the three covered by
$L_{\rm o}$ and $L_{\rm x}$).  In any case, the presence of such a
trend is indicative of the fact that through X--ray selection we
detect objects with a rather limited range in optical luminosity but a
large span in the radio one, extending the range of sampled $L_{\rm
r}$ towards faint sources compared to radio--selected quasars (see
Sect.~\ref{comp}).

\begin{figure}
\psfig{file=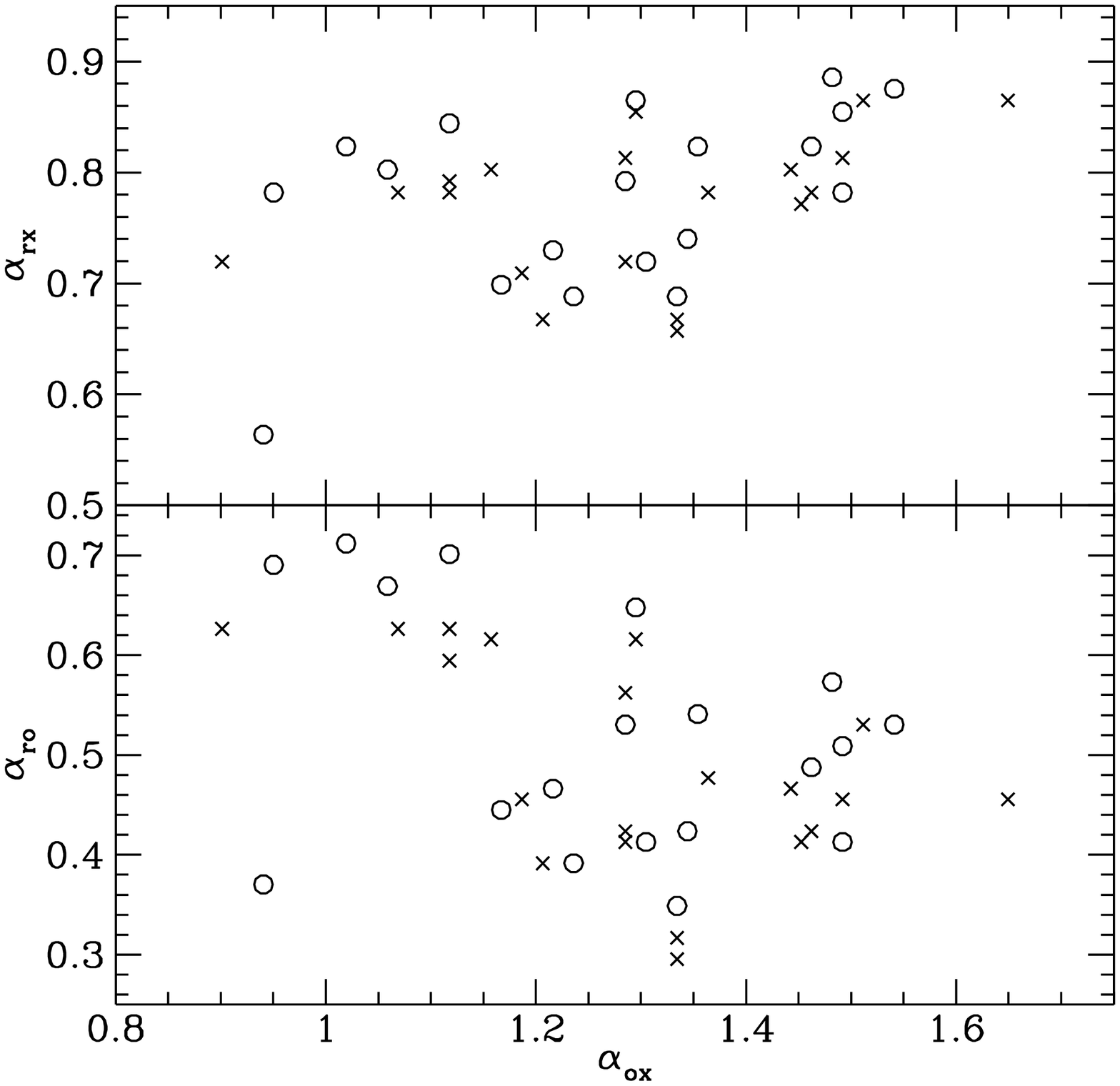,width=8truecm}
\caption{Broad band spectral indices: $\arx$ (upper panel) and
$\aro$ (lower panel) vs. $\aox$ for the ERL. Symbols
like in Fig.~\ref{figalr}}
\label{figalphas}
\end{figure}

On the other hand, it is interesting to consider the relative behavior
of the broad band spectral indices, as shown in Fig.~\ref{figalphas}.
$\arx$, which spans the smallest interval, and $\aox$, which spans the
largest one, do not show a strong correlation (98\% level, according
to the Student's-t test), while (a part for a conspicuous object, the
weak lined MS0815.7+5233 -- see also Sect. 4) $\aro$ decreases for
increasing $\aox$ (at $>$ 99\% level).  Interestingly, this trend is
qualitatively consistent with a change in the frequency position of
the peak of the synchrotron spectral component.  Furthermore, by
comparing Figs.~\ref{figalr} and ~\ref{figalphas}, it is clear that
this spectral behavior is connected with the source radio luminosity,
as the flatter is $\aro$, the lower is $L_{\rm r}$. Notice also that
the trend does not simply relate to the optical and X--ray
luminosities, which therefore appear to be worse tracers of systematic
behaviors in the SED shape.

We conclude that: a) the ERL sample (and the FS and SS subsamples)
significantly extends the radio luminosity range (but not the optical
one) of radio--loud quasars; b) the broad band spectral indices
behavior is consistent with being caused by a change in the
synchrotron peak energy; c) the trend in the SED shape correlates with
the radio luminosity.

\section{Comparison with other samples}
\label{comp}

We now compare the spectral properties of the ERL sample with those of
radio--selected quasars (FSRQ and Steep Spectrum radio--loud quasars, 
SSQ) and (radio and X--ray selected) BL Lac objects and
examine our findings in the light of the blazar spectral sequence
scenario already described and the unification of powerful radio
sources. Does the X--ray selection allow us to discover powerful
lined sources whose synchrotron emission peaks at high (UV-X--rays)
energies, analogously to X--ray selected BL Lac objects? Does the
detection of ERL force us to re--consider the above scenario?  Do the
nuclear properties of steep spectrum objects differ from those of the
flat spectrum sources or their nuclear properties are independent of
the extended emission?

Let us first examine the former issue.

The plane $\aox$ versus $\aro$ has been often used to characterize the
SED of the above classes of sources.  As mentioned in the
introduction, the position of the different objects in this diagram
can be accounted for by the different location of the synchrotron peak
in their energy distribution.

\begin{figure} 
\psfig{file=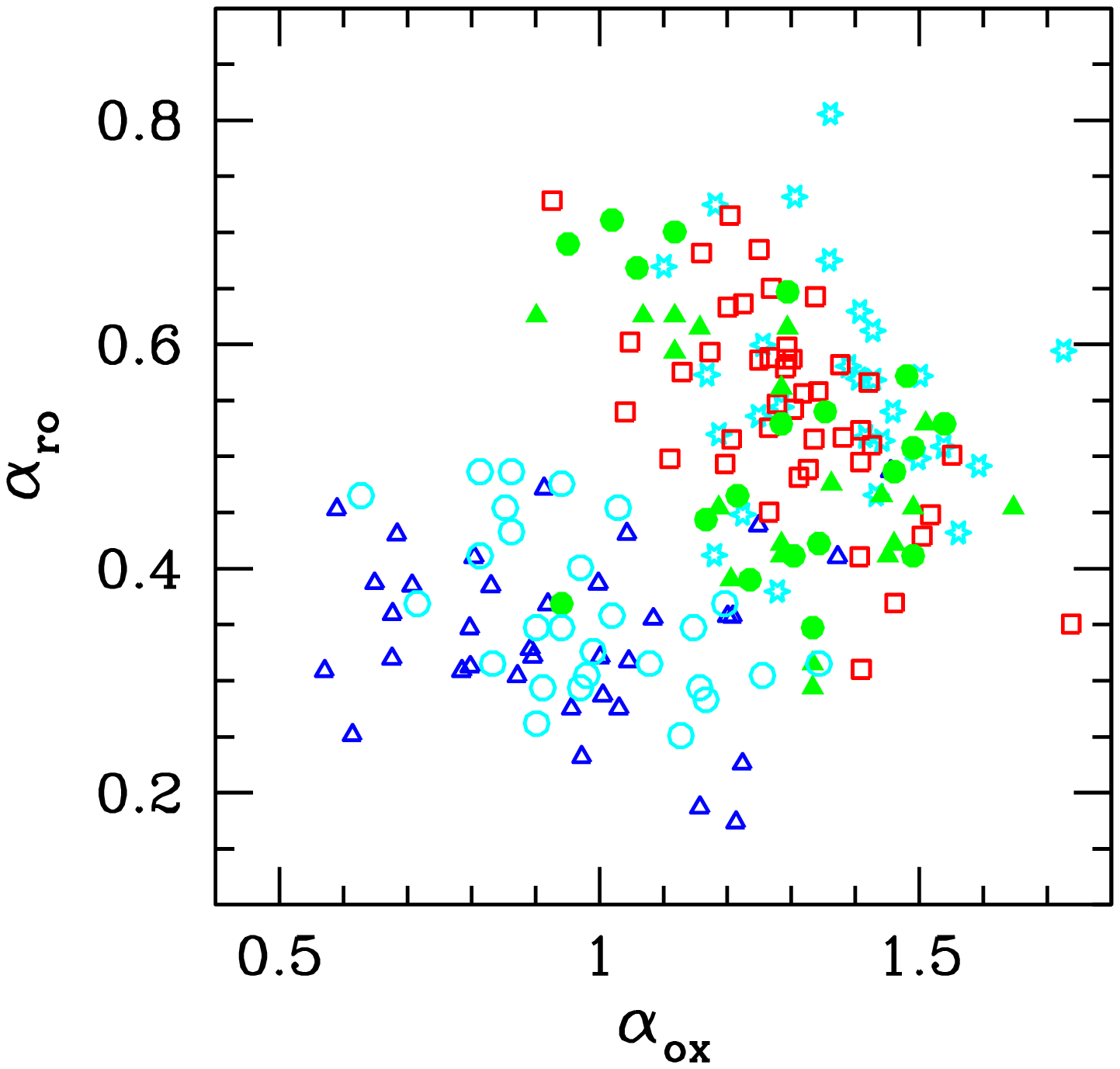,width=8truecm}
\psfig{file=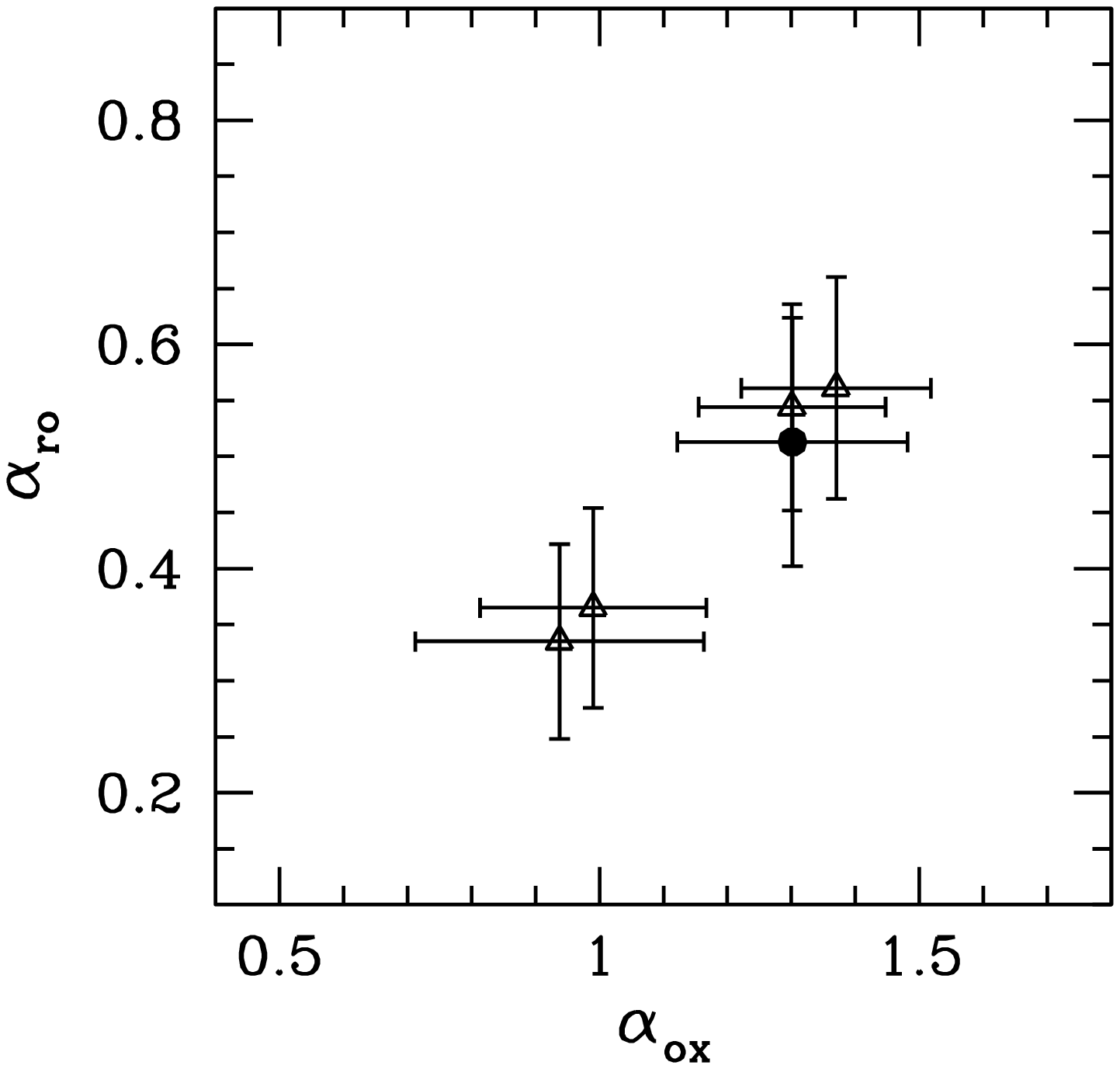,width=8truecm}
\caption{{\bf a)} Broad band spectral index $\aro$ vs. $\aox$ for the ERL
(filled triangles and circles for FS and SS, respectively) compared
to: EMSS BL Lacs (empty circles), Slew BL Lacs (empty triangles), 1Jy
BL Lacs (empty stars), 2Jy FSRQ (empty squares).  Just for graphical
purposes, we have not included in the plot the most recent blazars
selected by Perlman et al. (1998) and Laurent-Muehleisen et
al. (1999).
{\bf b)} The lower panel shows the loci of the different samples, the
black circle indicates the ERL sample, while the triangles the BL Lac
and FSRQ samples. The errorbars represent the width (1$\sigma$) of 
the distributions in $\aro$ and $\aox$ respectively. } 
\label{alp_all}
\end{figure}

In Fig.~\ref{alp_all}a we show the ERL and other samples (see the
figure caption) of both FSRQ and radio and X--ray selected BL Lacs.
For clarity, we also plot in Fig~\ref{alp_all}b a schematic representation 
of the loci occupied by the different samples, in which the mean and the 
1$\sigma$ width of the distribution in the two variables are plotted. 
The ERL occupy the transition region (in the ``boomerang'' shaped
blazar distribution) between FSRQ and X--ray selected ($\sim$ HBL) BL
Lac objects, roughly overlapping with radio--selected ($\sim$ LBL)
ones.  Note that the location of LBL and FSRQ in the spectral index
plane is due to the dominance in the X--ray band of a flat Compton
component.

Within the spread of the above quantities, this indicates that even
X--ray selection is unable to detect radio--loud broad--lined objects
with high peaked synchrotron components. Intriguingly, the object in
the left-bottom corner, in the ``HBL region'', is the
already mentioned MS0815.7+5233, a weak lined AGN, BL Lac-like (see
Stocke et al. 1991, Tables 11 \& 13 and notes therein).

Furthermore, the broad band spectral indices $\arx$ and $\aro$ of the
ERL tend to be lower ($\Delta \alpha \sim 0.1-0.2$) than those of the
FSRQ, with the flattest sources coinciding with the very same objects
which have lower $L_{\rm r}$. In other words ERL resemble radio
selected quasars in the optical and X--ray bands, while are
comparatively weaker in the radio one, indeed suggesting a larger
contribution (peak) to the SED at higher energies.  This result
is fully consistent with the high energy/X--ray selection of 
the sample.

\begin{figure}
\psfig{file=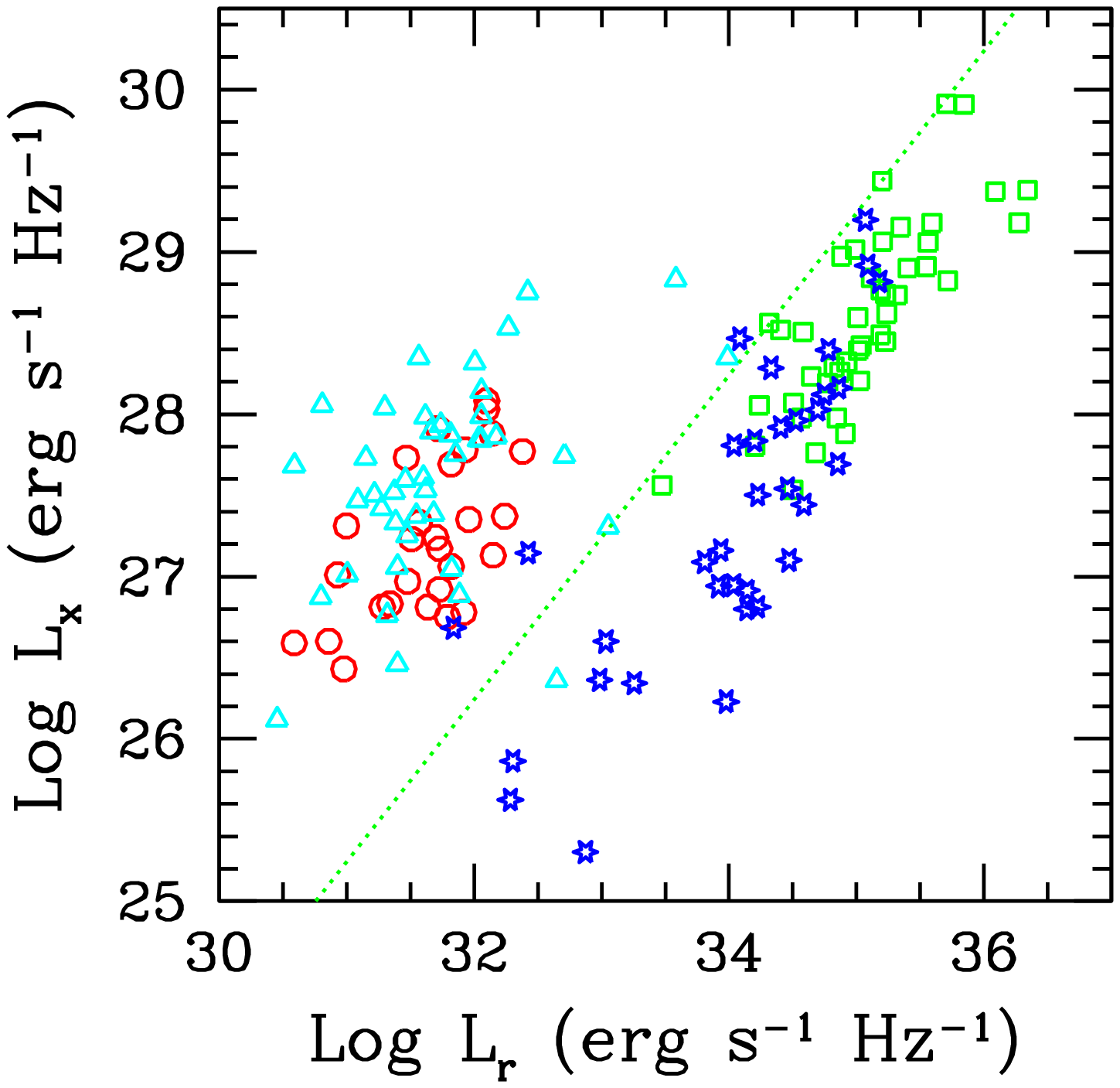,width=10truecm,height=8truecm,bbllx=0mm,bblly=65mm,bburx=200mm,bbury=220mm,clip=}
\caption{Monochromatic (@1 keV) X--ray vs monochromatic (@5 GHz) radio
luminosity of the ERL, EMSS BL Lacs, Slew BL Lacs, 1Jy BL Lacs, 2Jy
FSRQ (symbols as in Fig.~\ref{alp_all}). The dotted line indicates the
``nominal separation'' between HBL and LBL in the luminosity plane (as
defined in Fossati et al. 1998). }
\label{lrlx}
\end{figure}

It is also interesting to explore the relationship among the
luminosities of sources belonging to different samples, as the lack of
ERL with SED resembling those of HBL might be due, within the above
scenario, to their relatively higher luminosity.  In Fig.~\ref{lrlx}
the above samples are plotted in the $L_{\rm x}$ vs $L_{\rm r}$ plane
(luminosities uncorrected for evolution). While in the X--ray band the
ERL are on average a factor of  $\sim$ 10 weaker
than the FSRQ and similarly the SSQ, in the radio (core) luminosities
they are of the order of 100 (10)
times weaker than the FSRQ (SSQ) (e.g. Fossati et al. 1998; Maraschi
\& Rovetti 1994).  The comparison also shows that the $L_{\rm x}$ of
ERL covers the same range spun by HBL, although it extends at even
higher $L_{\rm x}$. Nevertheless there is no
indication of a trend for the weakest ERL to have SEDs more resembling
highly peaked sources, as their radio luminosity range is above
that of HBL.  We stress that in the very same EMSS sample HBL are
indeed found in large number (Morris et al. 1991, Wolter et al. 1991),
excluding observational biases against detection of ERL peaking in the
X--ray band.

These findings are therefore in global agreement with the expectation
of the blazars sequence scenario described in the introduction, thus
re-enforcing the view of a strong connection between the spectral
energy distribution and the radio luminosity and suggesting again that
the latter might be a good indicator of the source power.

However, it should be noted that ERL and
LBL span a similar range of three decades in $L_{\rm r}$.  While one
would expect some overlapping along the sequence -- as also suggested
by the existence of sources changing from being BL Lac--like to
quasar--like and vice versa, and RBL with FR~II extended morphology --
the extent of the common luminosity range seems to suggest more of a
parallel behavior, rather than a continuous trend, of lined and
lineless sources.  However, it appears that at lower $L_{\rm x}$
the ERL might be filling the intermediate region between LBL and
HBL. These two facts, and thus the overall distribution of radio and
X--ray luminosities of ERL, might be be ascribed to the increasing
dominance of a quasi--thermal optically--UV component in quasars of
increasingly lower (non--thermal) radio power. Indeed the typical
$\aro$ and $\aox$ of the low radio power ERL (see Fig.~\ref{figalr})
corresponds to SED whose peak is located in the optical--UV region.
The presence of a quasi-thermal component would indeed flatten $\aro$
and steepen $\arx$ in sources with decreasing non--thermal continuum.
This `excess' contribution could be identified with both/either `blue
bump' emission in quasars (FS) and/or contamination from the host
galaxy in SS (possibly Broad Line radio galaxies).

\section{Statistical and evolutionary properties}

Let us now examine the statistical properties of the sample within the
framework of the unification scheme for powerful radio sources. 
More specifically, do the steep radio sources evolve consistently with
the postulate of the unification scheme (i.e. similarly to flat
sources)?

\subsection{Number counts}

The number counts of radio--loud AGN of the EMSS have been presented
by Della Ceca et al. (1994).  Here we consider any possible
distinction related to the radio spectral index (FS vs SS), and
compare the ERL counts to those of other samples of radio--loud
objects.

\begin{figure}
\psfig{file=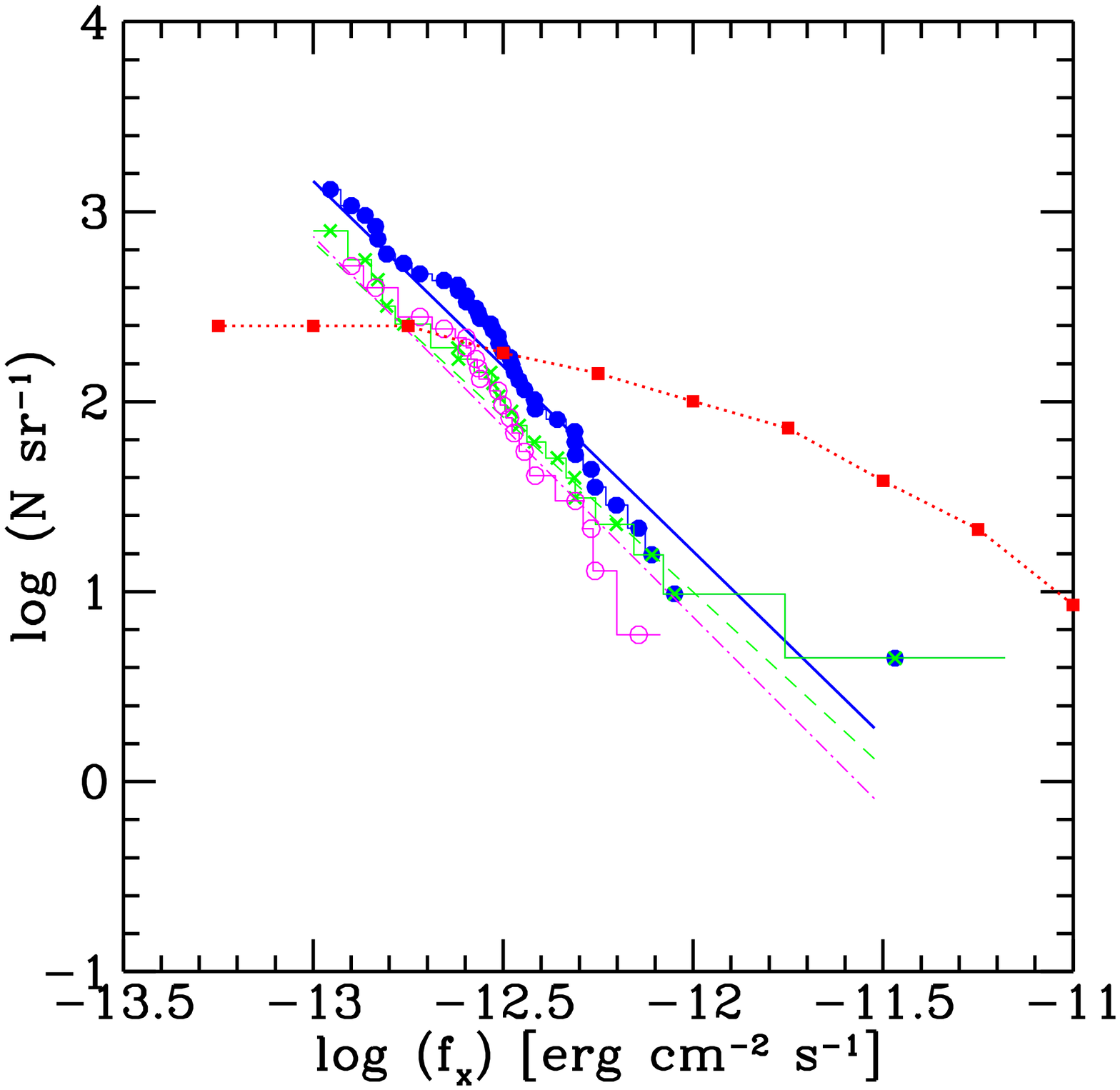,width=8truecm,height=7truecm} 
\caption{Integral number counts for FS (crosses), SS (circles) and all
ERL (filled circles). The lines represent the corresponding linear
fits: heavy line for ERL, dashed line for FS, dot-dashed line for SS.
The flatter distribution (squares) represents the EMSS BL Lacs (from
Wolter et al. 1991). Results of the fits are reported in
Table~\ref{stat}. }
\label{figlnls}
\end{figure}

The (integral) $\log N - \log S$ distributions for FS, SS, all ERL and
BL Lacs are shown in Fig.~\ref{figlnls}, and the parametric results
derived from their linear fits are reported in Table~\ref{stat}.  (The
fit to the counts is done by using the individual points in a
differential form, see e.g. Gioia et al. 1991 for a description of the
procedure).

\begin{table*}
\begin{center}
\caption{Statistical properties}
\begin{tabular}[h]{| l c c c  |}
\hline

      & ERL&  FS & SS \\

Count slope $^a$&1.9 [1.7-2.2]  & 1.8 [1.6-2.1] & 2.0 [1.7-2.3] \\

$\langle$ z $^b\rangle$ & 0.92 $\pm$ 0.40& 0.99 $\pm$ 0.40& 0.85 $\pm$ 0.39\\

$V_{\rm e}/V_{\rm a}$   & 0.78 $\pm$ 0.05 & 0.76 $\pm$ 0.06  & 0.81 $\pm$ 0.07 \\

C &   & 6.6 [5.6-7.4]  & 6.4 [5.6-7.1] \\

$\gamma$ & & 2.9 [2.5-3.1] & 3.0 [2.7-3.2] \\

LF$^a$ & 1.6 [1.4-1.7] & 1.6 [1.3-1.8] & 1.9 [1.6-2.2]\\

logK$^c$ & --8.74  & --8.95 & --9.00\\

\hline
\end{tabular}
\begin{list}{}{}
\item In square brackets the 1 $\sigma$ confidence range
\item $^{\rm a}$ Power-law index of a linear fit (integral slope)
\item $^{\rm b}$ $\pm$ 1$\sigma$ values
\item $^{\rm c}$ $K$ is in units of Mpc$^{-3} L_{44}^{LF-1}$,  where
$L_{44}$ is the luminosity in units of $10^{44} \ergs$.
\end{list}
\label{stat}

\end{center}
\end{table*}

The first point to be noticed is the  marginal (though it extends over
the whole range in flux) evidence for flatter counts of FS, which
dominate at the highest flux levels. The count distributions of FS is
in complete agreement with what found for all AGN -- mostly radio
quiet-- in the EMSS ($1.61 \pm 0.06$, Della Ceca et al. 1992), while
the slopes of SS and all ERL are marginally steeper (but only at 1
$\sigma$ level) than those of all AGN.  Clearly the statistics does
not allow us to perform more complex fits, although a single power law
might be inadequate to represent the distributions.

How do the ERL counts compare with those of FSRQ and SSQ?  Padovani \&
Urry (1992, hereafter PU92) examined the statistical properties of the
best defined radio sample, namely the 2 Jy (Wall \& Peacock
1985). This comprises 34 SSQ and 50 FSRQ
\footnote{We note here that the average radio spectral indices of the
two subsamples of FS and SS are fully consistent with those of FSRQ
and SSQ once the same definition is adopted (more precisely we obtain:
FS: 0.32 $\pm$ 0.36 (0.15$\pm$ 0.34), SS: 0.96 $\pm$ 0.14 (0.85 $\pm$
0.20), for a dividing $\alpha_{\rm r} = 0.7 (0.5)$}. 
The (poorly defined) count distribution in the radio
band is consistent with being Euclidean and also with those of FS and
SS (at 90\%) and of ERL (at 95\%) in the X--ray band.

The most interesting aspect is the extension to a factor 100 lower
fluxes provided by the X--ray selection of the ERL.  Indeed assuming
typical $\arx$ ($\sim 0.8$), $\langle \alpha_{\rm r} \rangle$ and
$\langle \alpha_{\rm x} \rangle $, an X--ray flux level of $\sim 10^{-12}
\ecs $ corresponds to $\sim 0.3$ Jy in the radio band (both for FS and
SS).  At these levels, the counts of FS exceed the prediction of the
beaming model proposed by PU92. Furthermore if one extrapolates to
lower radio fluxes (say by a factor $\sim 10$ as in the X--ray band),
the radio counts of X--ray selected objects are marginally consistent
with the Euclidean extrapolation from higher fluxes and thus largely
exceed (factor $>$10 for FS and $\sim$3 for SS) those predicted by the
beaming model, which drop below $\sim$ 0.1 Jy. Note that because of
the X--ray threshold the radio counts of ERL can be only a lower limit
to the actual surface density of radio--loud quasars.

It is also relevant to compare the ERL counts with those of BL Lacs
extracted from the EMSS. As shown in Fig.~7, the BL Lac distribution
is flatter, dominating by a factor $\sim 10$ at $\sim 10^{-12} \ecs$,
and an `inversion' of the two populations occurs just above $\sim$3
$\times 10^{-13} \ecs$.  Therefore while at high X-ray fluxes
RL quasars are as rare, or rarer, than BL Lacs, deep, small area
surveys are bound to find large numbers of them.

\subsection{Redshift distribution}

All sources are below redshift of $\sim 2$, with a broad peak around
$z\sim 1$. This is due to the sky coverage of the EMSS, combined with
the LF of the AGN. There is only marginal evidence for SS to be at
redshifts lower than FS (See Table~\ref{stat}. Further, the probability 
of getting by chance a larger KS value is 19.6\%).

The typical $z$ are globally similar (within 1 $\sigma$) to the
average redshifts of FSRQ and SSQ selected in the radio band (2Jy),
while they are significantly larger than those of FR~II radio
galaxies, as already noted by PU92.

\begin{figure}
\psfig{file=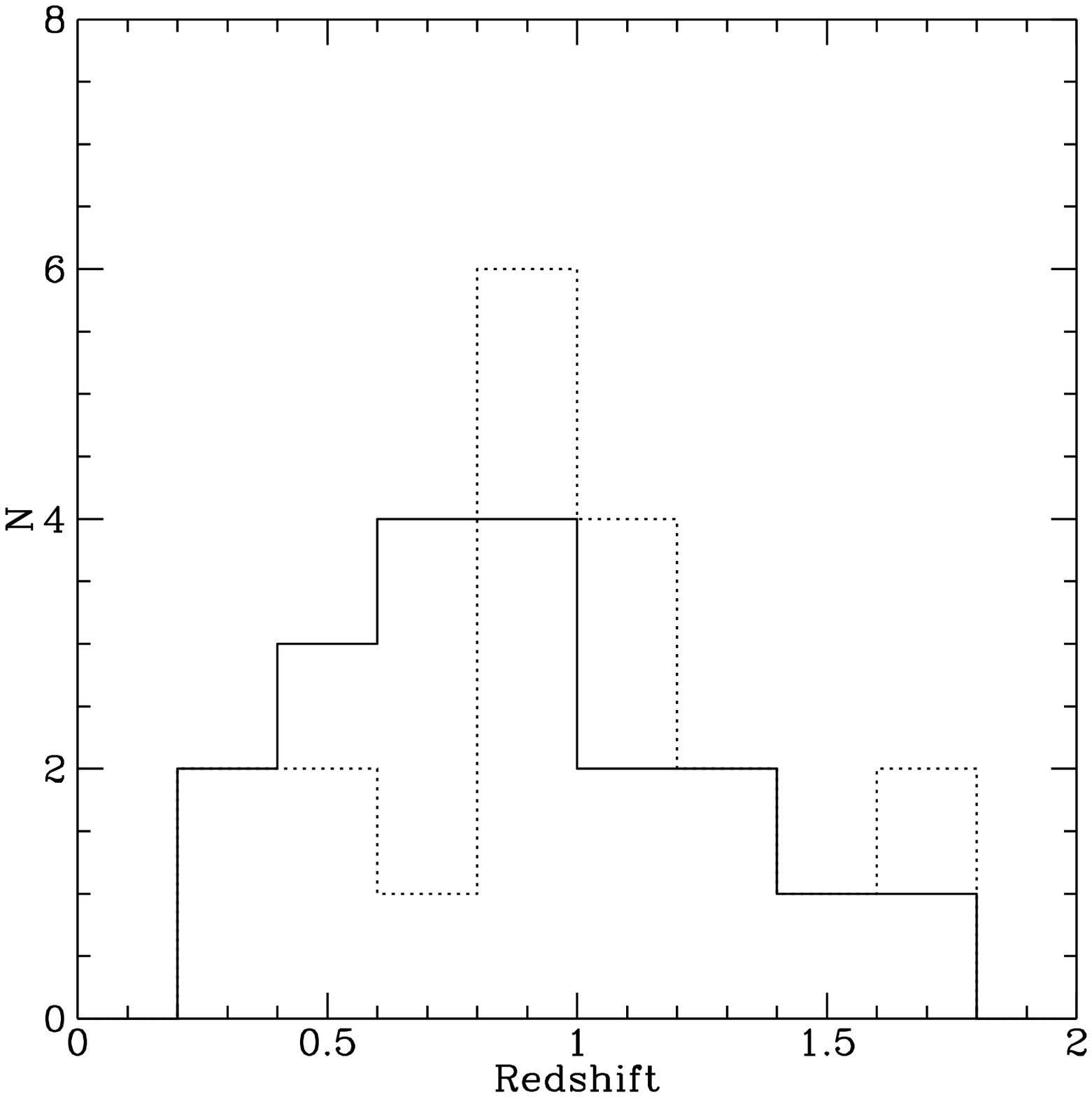,width=8truecm,height=6truecm}
\caption{Redshift distribution for SS sources (solid line) and FS
sources (dotted line).}
\label{figz}
\end{figure}

\subsection{$V_{\rm e}/V_{\rm a}$}

In order to quantify the evolutionary properties of the sample, we
estimated, again subdividing it into FS and SS, the $V_{\rm e}/V_{\rm
a}$ parameter (Avni \& Bachall, 1980), a generalization of the
$V/V_{\rm max}$ statistics (Schmidt 1968) when complete samples at
different flux limits are combined.  The results are shown in
Table~\ref{stat}.  Positive evolution is required for the subsamples
and the whole population at $> 4 \sigma$ level, while no statistical
difference can be found between the FS and SS behavior.

A systematically lower $V_{\rm e}/V_{\rm a}$ (though consistent within
$\sim 2 \sigma$) is found for the quasars of the 2 Jy sample (PU92),
possibly indicating a tendency of faster evolution of the X--ray
selected objects. \footnote{As this behavior is opposite to that of BL
Lacs derived from the EMSS, this indirectly (re)confirms the reality
of the uncommon negative evolution found for XBL.}

In order to compare the results with those relative to the 2Jy objects, 
we characterize
the evolution in terms of an exponential pure luminosity function of
the form $L(z) = L_{\rm (z=0)} e ^{-C\, t}$, where $1/C$ is the look--back
time in units of the Hubble time. Note however that for the whole AGN
sample in the EMSS, this form was found to be adequate only if a
luminosity dependent value for C was introduced (Della Ceca et
al. 1992).  The derived evolution rate is consistent (at $\sim$
2$\sigma$, see Table~\ref{stat}) with that inferred for FSRQ (and once
again it is indistinguishable between FS and SS). As our sample is
limited to $z<2$ nothing can be deduced on a possible decline in the
number density of sources at high redshifts (e.g. Dunlop \& Peacock
1990).

Also a power law luminosity evolution $L(z) = L_{\rm (z=0)}
(1+z)^{\gamma}$ can well represent the cosmological changes. The
required dependence on $z$ is similar to what is found for
radio--quiet objects (Table~\ref{stat}).  The results for the FS
sources are again consistent with the whole EMSS AGN behavior, while
SS require a slightly higher value of $\gamma$ (at 1$\sigma$).

For computing fluxes and luminosities we have assumed that
$\alpha_{\rm x} = 1$ -- supported by the fact that $\langle
\alpha_{\rm x} \rangle \sim 1$ in the ERL sample and by a study of the
IPC Hardness Ratio for the EMSS AGN (Maccacaro et al. 1988). Note
however that for X-ray spectra with a slope different from 1 the
effect of the cosmological evolution has the same functional form of
the K-correction factor and, in first approximation, the best-fit
value of $\gamma$ is lower (if $\alpha_{\rm x} > 1$) or higher (if
$\alpha_{\rm x} < 1$) by an amount $\sim |1 - \alpha_{\rm x}|$ (for
$|1 - \alpha_{\rm x}| < 1$).

\subsection{Luminosity function}

\begin{figure}
\psfig{file=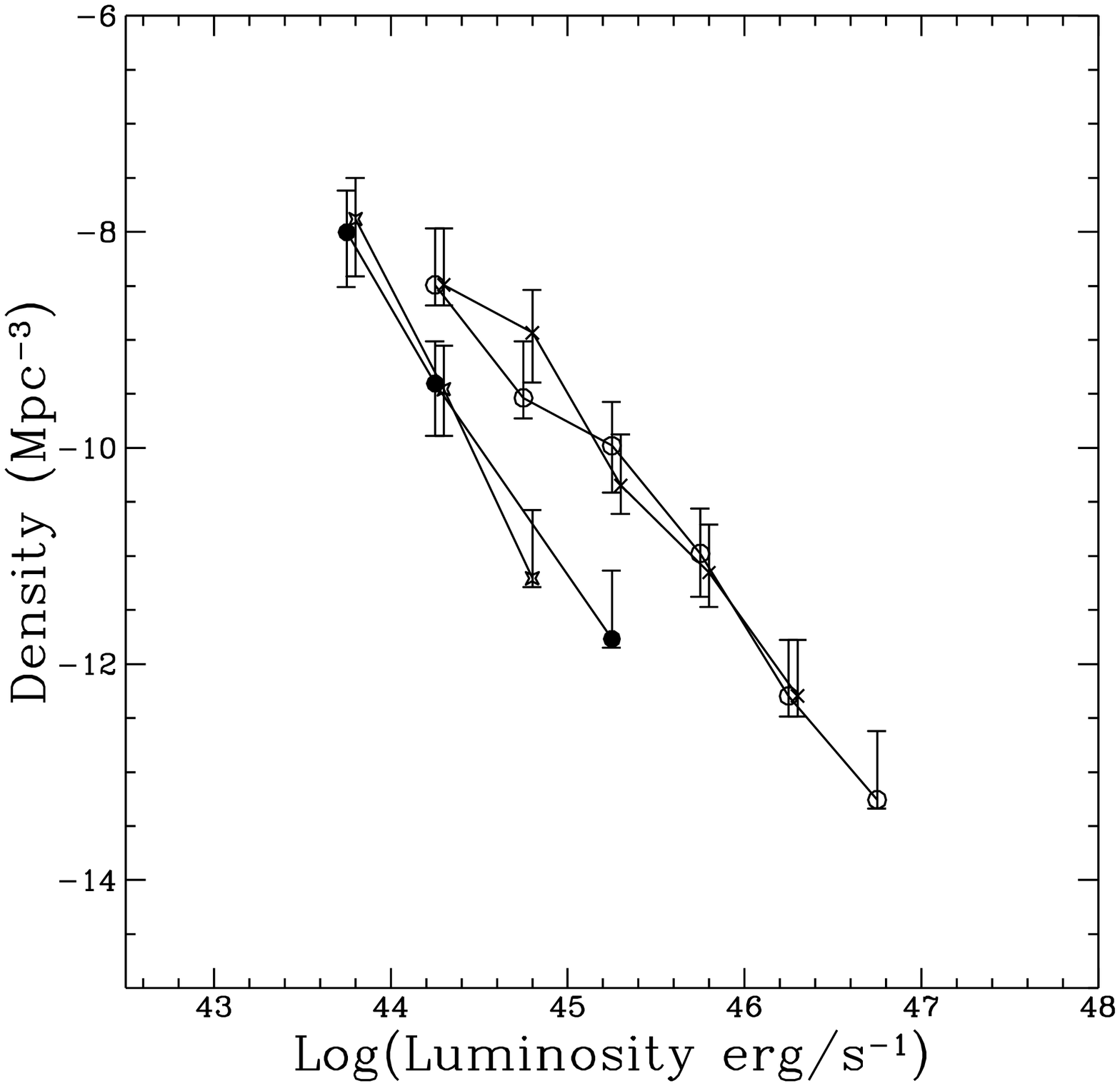,width=8truecm,height=6truecm}
\psfig{file=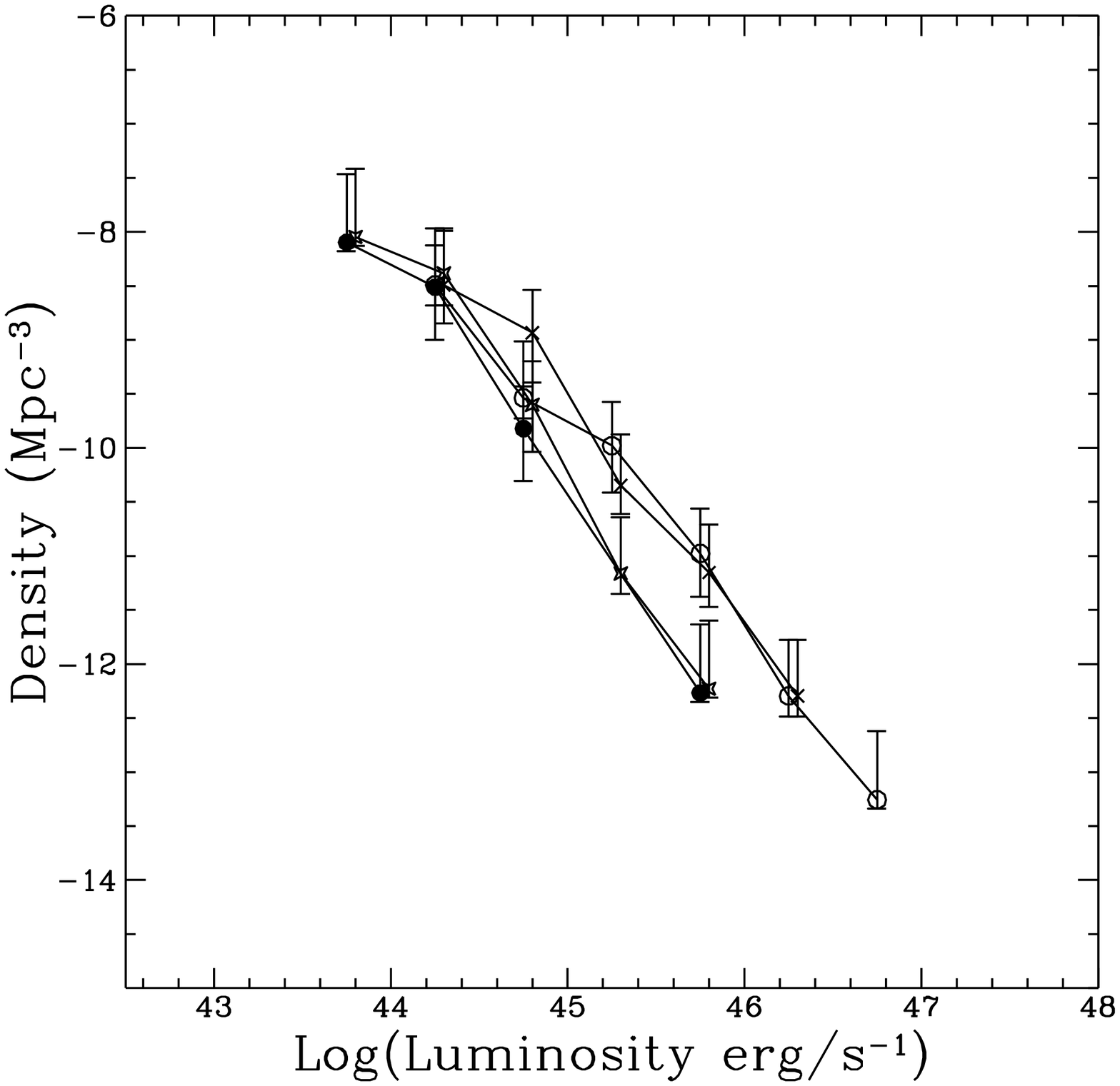,width=8truecm,height=6truecm}
\caption{Differential luminosity functions for the SS and FS sources,
in step of 0.5 $\log L$ (symbols as in previous figures; filled
circles and stars indicate de-evolved SS and FS sources
respectively). The SS points have been slightly shifted artificially
on the luminosity axis for clarity.  The de-evolving form is: {\it
upper panel} exponential, {\it lower panel} power law.}
\label{figlf1}
\end{figure}

Finally, we built the luminosity functions (LF) of FS and SS, both as
observed and applying the two evolutionary models to de-evolve
luminosities at $z=0$ (Figure~\ref{figlf1}a, b). As expected, 
no difference is found between flat and
steep spectrum objects (the extension of FS a factor of $\sim 10$ more
in luminosity is due to only one source). The distributions have been
fitted with power-laws (see results in Table~\ref{stat}), as the large
uncertainties do not allow to better characterize their shape.

At odds with these results, a significant difference in the number
density of flat and steep spectrum objects is found in radio selected
samples (2 Jy), with FSRQ numerically dominating above $\sim 10^{43}$
erg s$^{-1}$. Converted into the X--ray band, this luminosity
corresponds to $\sim 10^{45}$ erg s$^{-1}$, where however a smaller
number of SS, within the volume explored, are found.

The slope of the local LF of ERL is significantly steeper than that of
FSRQ, revealing a first clear difference between the population of
radio selected and X--ray selected flat spectrum quasars. This
discrepancy is of course present also with respect to the X--ray LF
computed by Maraschi \& Rovetti (1994) from the radio LF of FSRQ and
SSQ over the common range in X--ray luminosity.

The total number densities of FSRQ is predicted to be $\sim 2.2$
Gpc$^{-3}$ above $L_{\rm x}\sim 5\times 10^{43}$ erg s$^{-1}$ (PU92)
with a parent population of 81 FR~II Gpc$^{-3}$. Locally we find twice
this density.  Also, SS are $\sim 7$ times those predicted by
extrapolating below the cutoff the PU92's model predictions: in other
words we do not find evidence for either a cut off or a flattening of
the LF below X--ray powers of $\sim 3\times 10^{44}$ erg s$^{-1}$. XBL
have a slope slightly flatter than the ERL one and a local number
density (at say $10^{44}$ erg s$^{-1}$) about 50 times larger.

Also the ratio of FS/SS found in the EMSS significantly differ from
that estimated by Maraschi \& Rovetti (1994), i.e. a ratio of 30
between the two populations, while - at the corresponding X--ray
luminosity - we find similar numbers of FS and SS. In particular the
density of FS is about one tenth than their predictions.
\footnote{Note that for a flat spectral index $\alpha_{\rm r} \sim 0$ the
threshold flux would be comparable at higher selection radio
frequencies, therefore implying that the discrepancy is likely not due
to the (radio) selection band.}

The same `excess' of SS with respect to FS in the X--ray selected
sample clearly reflects in the relative ratio of SS/FS=19/20 found in
the EMSS, compared to 34/50 found in radio selected survey at higher
flux levels (2 Jy) -- despite SS do not dominate at low radio
luminosities.  This seem to imply that the radio selection is more
biased than the X--ray one in finding FSRQ with respect to SSQ.

\subsection{Predictions: number counts and redshift distributions} 

Finally, let us consider the number counts and redshift distributions
predicted by the above findings. In Fig.~10a the number counts
extrapolated at flux levels observable by Chandra and Newton-XMM are
reported. These have been derived by the integration of the evolving
LF (of FS and SS) up to redshift $z_{\rm max}\sim 5$ (between $L_{\rm
x} = 10^{41} - 10^{47}$ erg s$^{-1}$). The limits of the proposed
Newton-XMM surveys are plotted in Fig.~10a.  Given the areas that will
be probably covered at each flux limit ($\sim$ 1 , 5, and 150
sq. deg. respectively in the faint, medium and bright survey), we
expect about 2000 (300), 150 (50), 100 (45) SS (FS) quasars in each,
if the measured evolution is taken at face value and the extrapolation
holds at the faint fluxes.  These samples will be large enough to
allow refined prediction on the radio-loud quasars properties.  The
corresponding redshift distributions at different flux limits are
shown in Fig.~10b, where SS can be seen to dominate at lower $z$ at
all flux thresholds.

\begin{figure}
\psfig{file=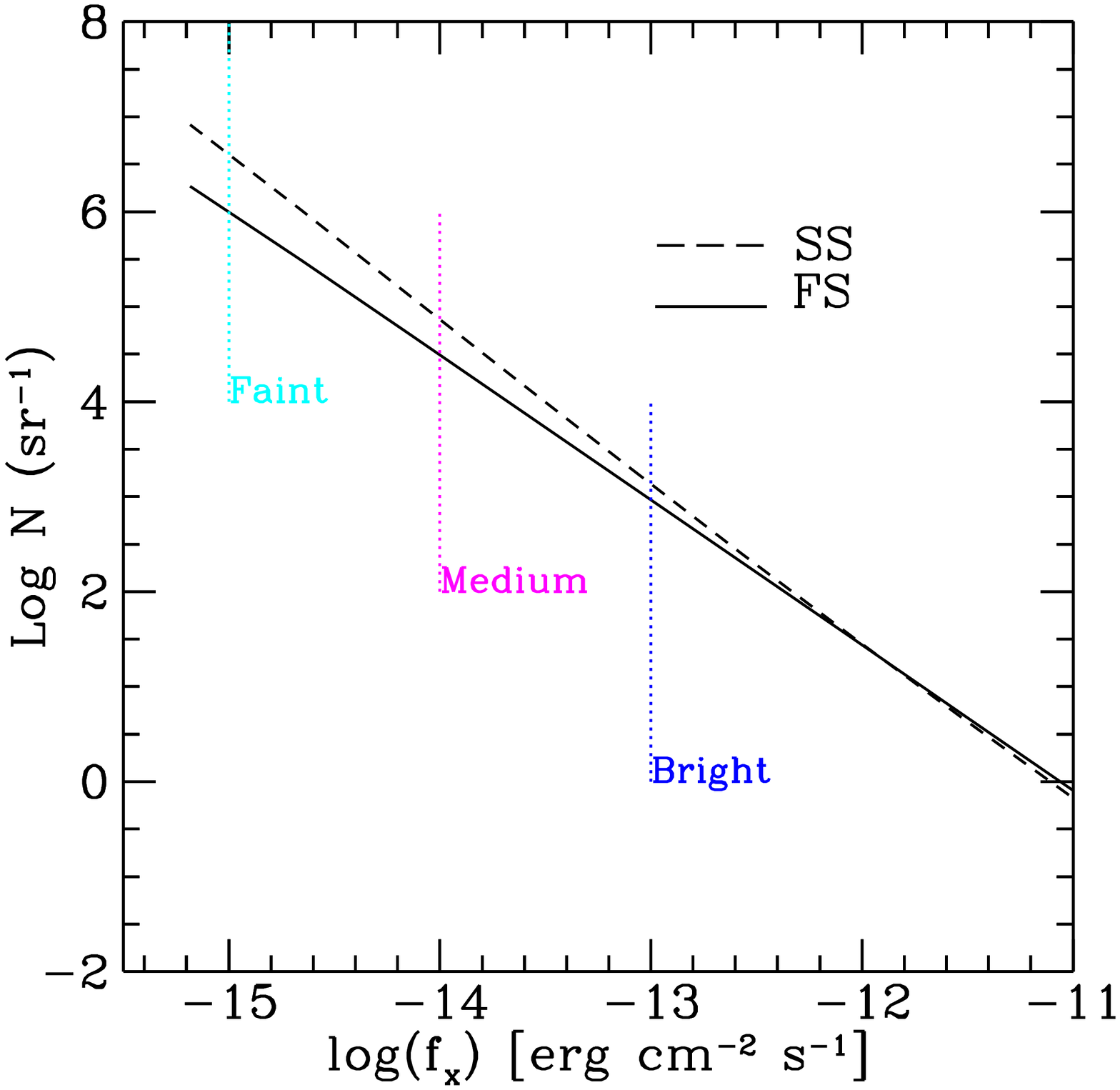,width=8truecm,height=6truecm}
\psfig{file=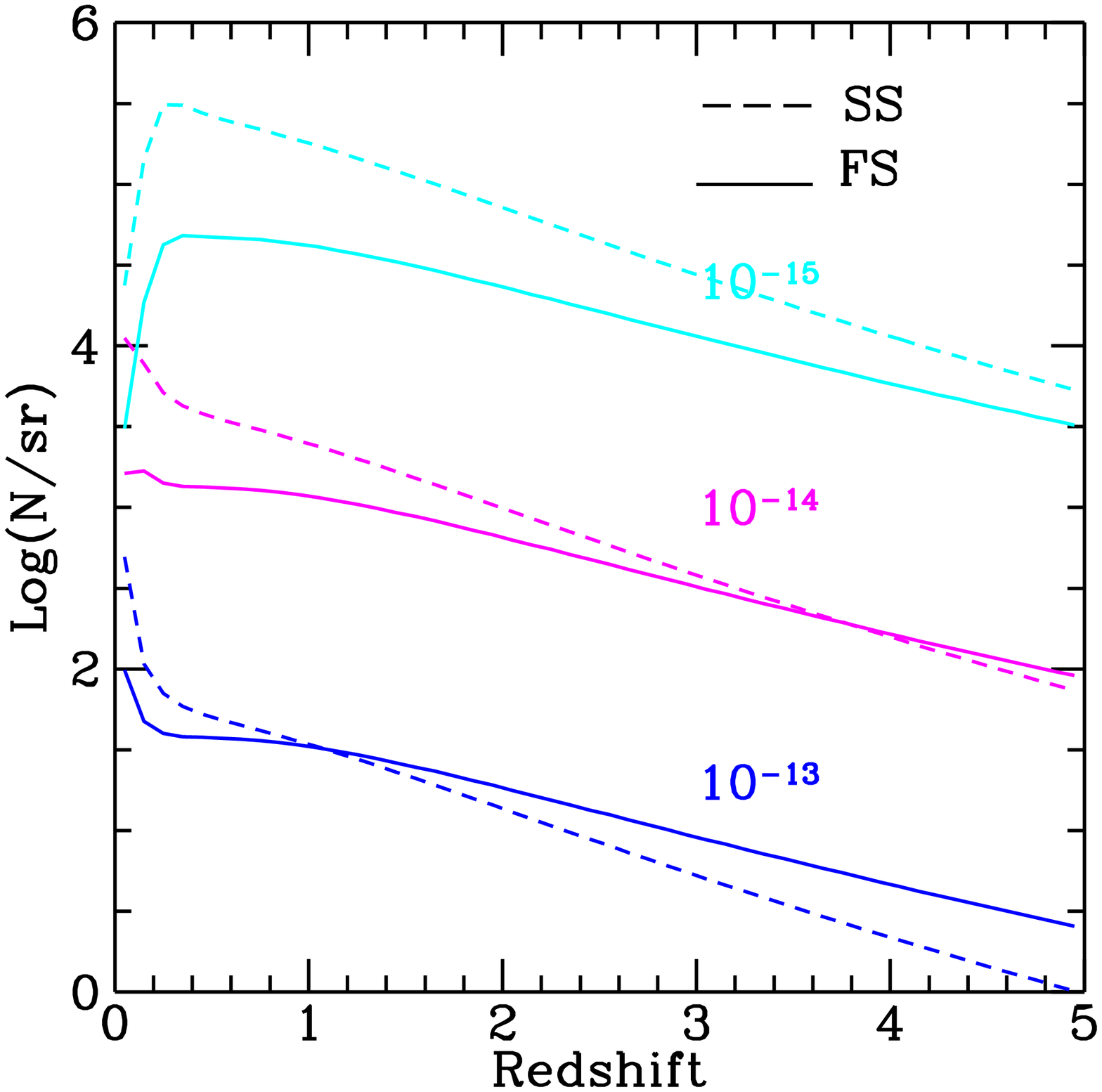,width=8truecm,height=6truecm}
\caption{a) Predicted number counts for the FS (solid line) and SS
(dashed line) at flux levels reachable by Chandra and Newton-XMM, in
the soft band, for a population extending up to $z_{\rm max} \sim 5$.
b) Redshift distributions of FS and SS, derived from the evolving LF,
at different flux limits. }
\label{figlf}
\end{figure}

\section{Summary and discussion} 

We have constructed the first complete X--ray selected sample of
radio--loud quasars.  This has allowed us to investigate both the spectral
properties and possible trends among blazars and the statistics of
such sources in the frame of unification models for radio--loud
objects.

Through X--ray selection (at the level of $\sim$ 10$^{-13}$ erg
cm$^{-1}$ s$^{-1}$) we explore the same range of redshifts and X--ray
and optical luminosities of radio--selected objects, but we find
quasars significantly fainter in the radio band.

As far as the spectral properties of blazars are concerned, FS ERL
do not differ significantly from radio selected sources.
However, the decrease in radio luminosity correlates with the
increasing frequency of the spectral peak, as indicated by the trend
of the broad band spectral indices. This supports the existence of a
sequence in the blazar family, where the peak decreases in energy with
increasing source radio power (Fossati et al. 1998; Ghisellini et
al. 1998).

Furthermore, although ERL are selected at X--ray luminosities
comparable to those of HBL, we find no evidence for the existence of
quasars whose synchrotron emission component peaks at X--ray
frequencies. This is also indicated by the similarity of $\arx$ of ERL
and FSRQ, which suggests that the synchrotron component does not 
dominate the X--ray band luminosity (see
e.g. Fossati et al. 1998; see also Padovani 2000).

On one hand in any unifying scenario some overlapping between objects
of different characteristics might be naturally expected (likely
corresponding to nuclei changing character or mismatching the extended
radio morphologies). On the other hand the ERL spectral distributions
appear to strongly resemble those of LBL over a large range of powers
($L_{\rm r}$ spans three decades), where both broad lined and lineless
objects co--exist. This suggests that the line typology might not be
univocally related to the source power and broad band spectral shape.
Indeed, one might envisage that the (occasional) appearance of broad
line components in BL Lacs is related to a change occurring in the
nuclear source -- but at such low luminosity levels that the lines
never become quasar--like -- while the blazar sequence of power vs
continuum spectral behavior would be a separate effect and the
resemblance of the broad band spectral indices of LBL and ERL might be
indeed attributed to a similar trend of the synchrotron peak position
with (radio) luminosity.

However, from the paucity of information we have on ERL, this
similarity can be due to the progressive dominance of a broad thermal
component (blue bump or contribution from the host galaxy in FS and
SS, respectively) over the non--thermal emission in objects of lower
non--thermal radio luminosity.  Accounting for this effect in fact
reduces the intrinsic overlapping of the two populations: decreasing
the optical luminosity by considering only the (actual) non--thermal
emission corresponds to move ERL towards FSRQ in the spectral index
plane.  If so, the tighter relation between source power and SED shape
(and line strength) would be re--enforced by the ERL sample.  Clearly
this ambiguity can be solved by the estimate of the optical and X--ray
spectra of ERL. Support to this view also comes from the findings of
the REX survey (Caccianiga et al. 2000), where a significant number of
sources are found with clear optical excesses with respect to FSRQ and
LBL (see their Fig.~6), while the radio and X--ray luminosities are
simply linearly correlated.

We stress here that also the strong difference in the statistical
properties of lineless and lined samples indicates that the two
populations are not simply analogous manifestations of the same
phenomenon, where only the line intensities are different.

We find no indication of any dependence of the spectral properties on
redshift, nor any difference between the nuclear properties of flat
and steep spectrum radio quasars, when selected in the X--ray band,
except possibly flatter counts for the FS. In other words the nuclear
component seems not to be directly dependent on the extended (radio)
properties. 

From the statistical point of view, the radio and X--ray selection
sample the same population at the brightest fluxes.  The main new
information we obtain is that at the lower flux levels sampled through
the X--ray selection the radio source counts do not drop -- and not
even flatten -- as predicted by some beaming models (e.g. PU92). 
Surveys at low radio flux levels ($\sim$ mJy) will be able to
establish the presence or absence of such a decline in the
population.

The ERL sample has provided also with the possibility of directly
comparing the X--ray evolution of quasars with that of BL Lacs. This is
particularly relevant both for understanding the reason of the
negative evolution of BL Lacs in this band and in particular as a
direct test of models postulating a physical evolution of radio--loud
quasars into BL Lacs (e.g. Cavaliere \& Malquori 1999).

The ERL population significantly evolves, at a rate indistinguishable
from that of FSRQ. With respect to the radio selection, the X--ray one
provides also with a different number ratio of flat and steep radio
sources, with the latter comparable to the former ones. This number
ratio does not appear to depend on the radio luminosity, but
significantly depends on the radio -- and possibly on the X-ray --
flux, being the X--ray selection less biased at low radio fluxes. Both
deeper radio surveys and the definition and study of optically
selected blazars -- as that which will be provided by the EIS surveys
(e.g. Cagnoni \& Celotti 2001) -- should confirm this trend.

Due to the dramatic effect of beaming in altering the shape of the
luminosity functions, it is not possible to simply use the above ratio
as an indicator of the beaming angles subtended by FS and SS. Recent
indications of the existence of a slower moving `layer' jet component
dominating the observed emission in radio galaxies (e.g. Laing 1993,
Chiaberge et al. 2000) further modify the expected number ratio of
beamed and parent population sources.  Furthermore as the luminosity
functions do not show a turn over at low powers it is not even
possible to provide model independent estimates on the total number of
objects of the two populations.  Nevertheless it is clear that the
beaming parameters and number density of the parent and beamed
populations have to be revised in the light of the new and less biased
information provided by the ERL, affecting the predictions on the
radio-activity in AGN and galaxies as well as the contribution of
these objects to the $\gamma$--ray background.

\appendix
\section{Sample selection: notes on individual objects}

Two of the 43 radio--loud objects of the EMSS have been classified as
Narrow Emission Line objects (MS1252.4-0457 as a Seyfert 2 galaxy, and
MS1414.8-1247 as an HII region; Boyle et al. 1995) and therefore we
remove them from the sample under study.

MS0234.8+0655 is detected at 6cm, but not at 20cm. By using the EMSS
radio flux the source is radio--loud ($\aro$=0.43), but with 
the NVSS flux limit (= 2.5 mJy) $\alpha_{\rm r} \leq -0.81$.  Therefore
both MS0234.8+0655 and MS2134.0+0028 ($\alpha_{\rm r} = -0.88$) have a
highly inverted spectrum and are removed from the sample.

We examined also the case of MS1253.6-0539 that has a very high upper
limit from the EMSS data (5.5 mJy at 5 GHz) and therefore in principle
could be a radio source. However, the object is also not detected at
1.4 GHz by the NVSS at a limit of 2.5 mJy.  Furthermore, it appears in
Table 10 of Stocke et al. (1991) as having an ambiguous spectrum due
to the poor signal to noise. We therefore do not include it in the
sample.

Note that MS0833.3+6523 (a.k.a. 3C204) has F$^{\rm 6cm}_{\rm core}
\sim 30$ mJy (from the EMSS VLA data) while F$_{\rm tot} = 369 $ mJy:
the extended flux dominates the emission and we use the latter to
compute the radio spectral index.  The same applies to MS1623.4+2712,
for which the EMSS F$^{\rm 6cm}_{\rm core} = 43$ mJy while F$_{\rm
tot}$ = 240 mJy.

\begin{acknowledgements}
We thank Alessandro Caccianiga, Roberto Della Ceca and Laura Maraschi 
for useful comments and discussions.  
The Italian MURST is acknowledged for financial
support - partly under the Program Cofinanziamento 1999.
\end{acknowledgements}

\end{document}